\documentclass[11pt]{article}

\usepackage[final]{acl}

\usepackage{times}
\usepackage{latexsym}

\usepackage[T1]{fontenc}

\usepackage[utf8]{inputenc}

\usepackage{microtype}

\usepackage{inconsolata}

\usepackage{graphicx}

\usepackage{subcaption}
\usepackage{algorithm}
\usepackage{algpseudocode}
\usepackage{enumitem}
\usepackage{tabularx}
\usepackage{threeparttable}
\usepackage{color, xspace}
\usepackage{appendix}
\usepackage{amsmath}
\usepackage{amssymb}
\usepackage{booktabs}        
\usepackage{multirow}        
\usepackage{array}           
\usepackage{caption}         
\usepackage{makecell}
\usepackage{graphicx}

\title{MemSearch-o1: Empowering Large Language Models with Reasoning-Aligned Memory Growth in Agentic Search}

\author{
 \textbf{Sheng Zhang\textsuperscript{1}},
 \textbf{Junyi Li\textsuperscript{1}},
 \textbf{Yingyi Zhang\textsuperscript{1,2}},
 \textbf{Pengyue Jia\textsuperscript{1}},
 \textbf{{Yichao Wang}\textsuperscript{3}\footnotemark[1]},
 \\
 \textbf{Xiaowei Qian\textsuperscript{1}},
 \textbf{Wenlin Zhang \textsuperscript{1}},
 \textbf{Maolin Wang\textsuperscript{1}},
 \textbf{Yong Liu\textsuperscript{3}},
 \textbf{Xiangyu Zhao\textsuperscript{1}\thanks{Corresponding authors: \href{mailto:wangyichao5@huawei.com}{wangyichao5@huawei.com} (Yichao Wang), \href{mailto:xianzhao@cityu.edu.hk}{xianzhao@cityu.edu.hk} (Xiangyu Zhao).}}
  \\
\\
 \textsuperscript{1}City University of Hong Kong,
 \textsuperscript{2}Dalian University of Technology,
 \textsuperscript{3}Huawei Technologies Ltd.,
\\
 \small{
   \textbf{Correspondence:} 
   \href{mailto:wangyichao5@huawei.com}{wangyichao5@huawei.com},
   \href{mailto:xianzhao@cityu.edu.hk}{xianzhao@cityu.edu.hk}
 }
}

\usepackage{color, xspace}

\begin{document}
\maketitle
\begin{abstract}
Recent advances in large language models (LLMs) have scaled the potential for reasoning and agentic search, wherein models autonomously plan, retrieve, and reason over external knowledge to answer complex queries. 
However, the iterative think–search loop accumulates long system memories, leading to memory dilution problem. In addition, existing memory management methods struggle to capture fine-grained semantic relations between queries and documents and often lose substantial information.
Therefore, we propose \textbf{MemSearch-o1}, an agentic search framework built on reasoning-aligned memory growth and retracing. MemSearch-o1 dynamically grows fine-grained memory fragments from memory seed tokens from the queries, then retraces and deeply refines the memory via a contribution function, and finally reorganizes a globally connected memory path. This shifts memory management from stream-like concatenation to structured, token-level growth with path-based reasoning. Experiments on eight benchmark datasets show that MemSearch-o1 substantially mitigates memory dilution, and more effectively activates the reasoning potential of diverse LLMs, establishing a solid foundation for memory-aware agentic intelligence.
Our code is available at \url{https://github.com/Applied-Machine-Learning-Lab/ACL2026_MemSearch-o1}.
\end{abstract}

\section{Introduction}
In recent years, retrieval-augmented generation (RAG)~\cite{RAG} has emerged as a powerful framework that enables large language models (LLMs) to access external corpora by retrieving text chunks relevant to a given query~\cite{lsrp, zhao2019deep}. While effective in enhancing LLMs with factual knowledge, RAG often provides shallow support: retrieved passages are limited in scope, and the pipeline lacks explicit reasoning over the original query~\cite{evoking}. This significantly constrains performance on complex, multi-hop problems. To overcome these limitations, the paradigm of deep search has been proposed~\cite{memory_survey, searcho1}. Unlike conventional RAG, deep search autonomously plans, retrieves, reflects, and reasons over external knowledge in an iterative manner, constructing deeper reasoning chains through repeated interactions with knowledge sources~\cite{searchr1, searcho1}. Deep search is especially well-suited to LLMs with advanced reasoning capabilities~\cite{wang2025tutorial, liu2024deepseek, yang2025qwen3, zhang2025deep, wu2026deepresearch}, which effectively exploits the complex knowledge~\cite{ferrag2025llm, liu2025large} and perform high-quality generation~\cite{liu2025llmemb, wang2025rethinking} based on the search trajectories.


Despite its promise, the deep search paradigm still faces two critical limitations.
First, the accumulated thinking history and redundant document fragments often introduce irrelevant information~\cite{memoryr1, Graphr1, graphragr1, wenmemory}, while attention dilution in LLMs causes much of the context to be overlooked~\cite{lost_middle, derongmem}.
As a result, key evidence may remain undiscovered, ultimately degrading reasoning quality. This issue becomes more severe as system memory grows: the signal-to-noise ratio declines, making it increasingly difficult for the model to focus on query goals.
Second, although advanced LLMs possess substantial latent reasoning capacity enabled by their scale~\cite{scaling_law}, current deep search studies have yet to fully exploit this potential~\cite{memory_survey}. In addition, since memory dilution cannot be solved by prompt engineering alone, effective explicit memory management strategies are essential to unlock the reasoning capabilities of LLMs.

Prior efforts to address memory dilution in deep search can be broadly divided into two paradigms. The first focuses on memory summarization and refinement, where retrieved documents are compressed~\cite{searcho1,mem0} or filtered~\cite{amber} to improve reasoning quality. 
While intuitive, these methods often fail to adapt to the evolving semantics of updated queries, resulting in the loss of critical information.
The second paradigm centers on memory pruning and distillation, where the system selectively retains or compresses past context for subsequent reasoning through designing special prompts~\cite{mem1,memoryr1}.
However, these approaches largely optimize for overall task rewards without explicitly modeling the structure and logic of stored memories, leading to suboptimal reasoning paths. Even methods that explicitly retrieve key elements from refined memories~\cite{amem} risk discarding important connections, as their quadratic search mechanisms may overlook latent semantic dependencies across memory fragments.



To overcome these challenges, we propose \textbf{MemSearch-o1}, a novel agentic search framework that emphasizes fine-grained, reasoning-aligned memory growth. 
Instead of refining memories based on complex query-document semantic relations, our approach expands memories by collecting contents associated with specific tokens of subject, action, degree, temporal~\cite{part_of_speech}, i.e., seed tokens of the query, thereby aligning more closely with the search goal. The framework then retraces and filters these token-associated memories to build a semantically coherent and logically structured memory path for answer generation.
By enabling the agent to operate within compact, query-focused memory spaces, rather than diluted long contexts, MemSearch-o1 provides a concise yet comprehensive semantic environment optimized for reasoning. This targeted memory management substantially enhances the depth, precision, and overall quality of the deep search.


Our main contributions are as follows:
\begin{itemize}[leftmargin=*]
\item We propose MemSearch-o1, the first deep search framework that grows fine-grained memory fragments from query tokens, enabling clearer semantic alignment and stronger reasoning.
\item We develop a retracing-based memory refinement mechanism that filters and reorganizes fragments into concise, coherent memory paths optimized for multi-hop reasoning.
\item Extensive experiments on eight QA benchmarks and LLMs show that MemSearch-o1 effectively mitigates memory dilution and consistently outperforms strong baselines.
\end{itemize}
\label{overview}
\begin{figure*}[t]
    \centering
    \includegraphics[width=\linewidth]{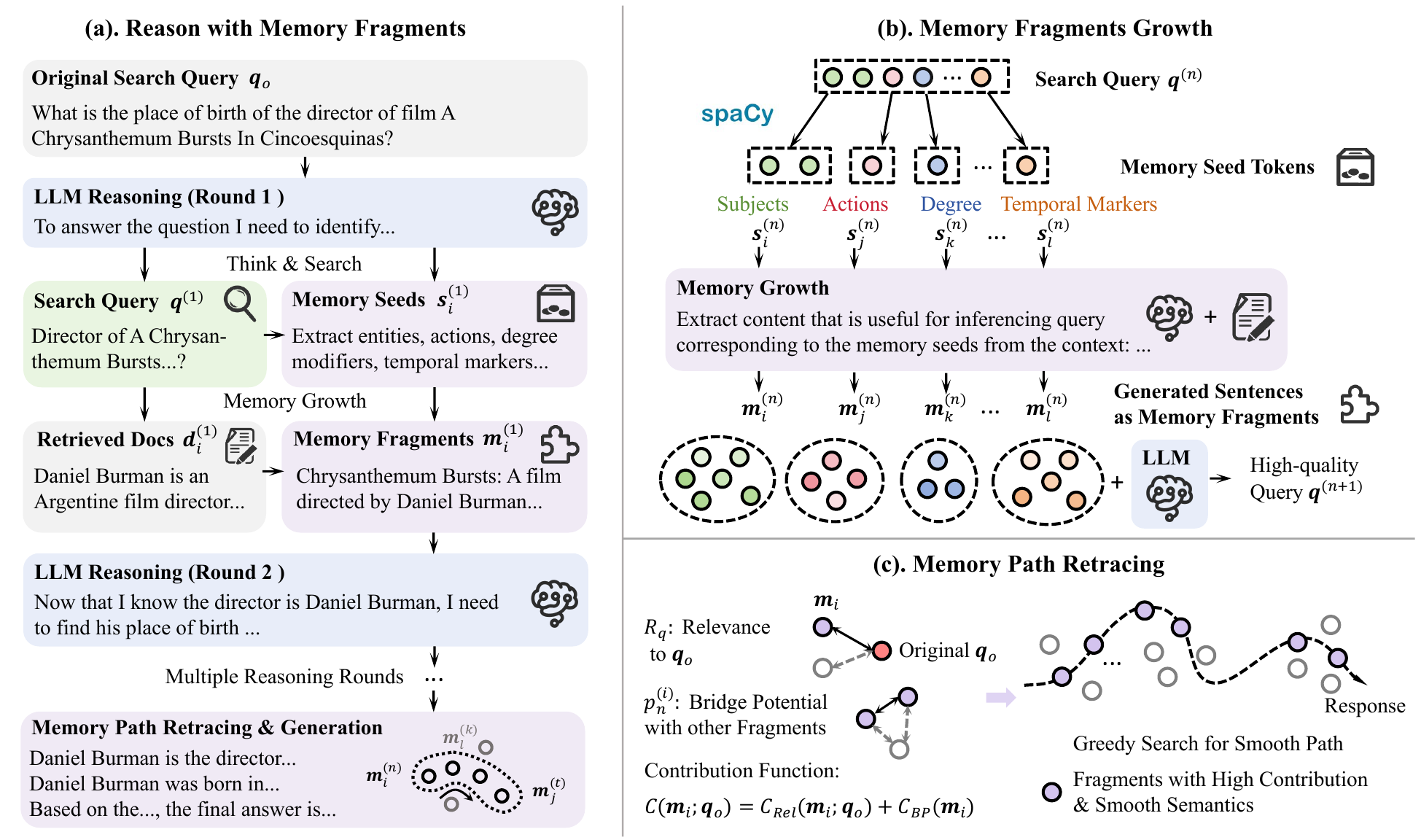}
    \caption{Overview Framework of MemSearch-o1. (a). Our MemSearch-o1 enables memory growth and retracing in the agentic. (b). The memory fragments are grown from the seed tokens extracted from the search queries, which enables efficient and effective semantic exploration. (c). Retrace and construct a memory path based on their relevance to the original question and the relevance among the memory fragments to enhance the reasoning process.}
    \label{framework}
\end{figure*}
\section{Method}
In this section, we will introduce the MemSearch-o1 in detail. We first give an overview of our framework (Section \ref{overview}), then introduce the memory seeds extraction from queries (Section \ref{seeds}), followed by memory fragments growth and memory path retracing (Section \ref{fragments} and \ref{manifold}).

\subsection{Overview}
\label{overview}
Most agentic search frameworks produce especially long working trajectories during the iterative reasoning process~\cite{gao2025navigate, peng2025stepwise}, leading to a serious memory dilution problem~\cite{memory_survey, lost_middle}. Although existing memory management methods can refine the redundant memories~\cite{autorefine, mem0}, they only focus on the semantic relevance to the query, which may either be too complex for LLMs to comprehend or lead to a dramatic information loss. To tackle this issue, we develop MemSearch-o1, an agentic search framework that employs reasoning-aligned memory growth and retrace for LLMs with reasoning ability. 

The overall framework is shown in Figure \ref{framework}. 
The MemSearch-o1 starts from an original search query $\boldsymbol{q}_o$, and a predefined instruction for LLM reasoning. Then, the LLM raises a new required query for search, and we utilize this query for \textbf{memory seeds preparation}. Specifically, the raised query will be split into memory seeds, and each seed consists of several memory seed tokens grouped by their parts of speech~\cite{1957linguistic, part_of_speech}. After that is the \textbf{memory fragments growth} stage, where relevant documents are retrieved using the latest raised query, and the memory seed tokens are fed into the LLM to guide the information extraction. The grown fragment sentences are then conveyed into the LLM again for next-round reasoning. The LLM reasoning round, together with corresponding memory growth, is iteratable. Finally, we collect all the fragment sentences for \textbf{memory path retracing} to deeply refine the memory. We design a contribution function to select fragments with high relevance to the original query and high bridge potential with other fragments. Finally, we use greedy search to find a semantically smooth memory path for the final generation. The whole framework improves the capability of our deep search system to exploit essential information for concise yet semantically rich memory growth and perform higher-quality reasoning and generation.

\subsection{Memory Seeds Preparation}
\label{seeds}
Most existing memory management approaches rely on LLMs to refine retrieved documents or to directly fetch relevant fragments from stored memories, but such strategies often incur significant information loss~\cite{infogain}. To tackle this issue, our method grows fine-grained memories from seed tokens extracted from queries, enabling the construction of a concise yet semantically rich memory space that remains closely aligned with the evolving query goals.


Inspired by linguistics, a sentence is composed of words with eight parts of speech, among which \emph{nouns, pronouns, verbs, adjectives}, and \emph{adverbs} usually contain the richest semantics~\cite{part_of_speech}.
In the context of deep search and memory growth, nouns and pronouns often denote \textbf{subjects} such as people, places, and objects, as well as \textbf{temporal markers} that indicate the time or duration of events. Verbs, by contrast, usually capture the \textbf{actions} performed by or involving these subjects, providing essential links within reasoning chains. Adjectives and adverbs function as \textbf{degree modifiers}, describing the properties of subjects or the manner and intensity of actions.
Based on these observations, we categorize nouns and pronouns into subjects and temporal markers, and group adjectives and adverbs together as degree modifiers. This decomposition serves as the foundation for our token-level memory growth strategy, enabling fine-grained alignment between the goals of queries and memory exploration.



Specifically, let $\boldsymbol{q}^{(n)}=[{q}_1^{(n)}, {q}_2^{(n)}, \dots, {q}_L^{(n)}]$ denote a search query consisting of $L$ tokens at the $n$-th turn. We then identify and partition \(\boldsymbol{q}^{(n)}\) into \(L_r\) memory seeds as follows:  
\begin{equation}
\mathcal{S}^{(n)} = \big\{ \boldsymbol{s}_1^{(n)}, \boldsymbol{s}_2^{(n)}, \dots, \boldsymbol{s}_{L_r}^{(n)} \big\}   \;\xleftarrow\; \boldsymbol{q}^{(n)},
\end{equation}  
where each seed \(\boldsymbol{s}_i^{(n)} = \big\{ {q}_j^{(n)} \big\}_{j=1}^{z_i}\) contains $z_i$ tokens from $\boldsymbol{q}^{(n)}$ and corresponds to one type from subjects, actions, degree modifiers, and temporal markers. 
For efficient implementation, we use the spaCy\footnote{https://spacy.io/} toolkit to identify and group these tokens into respective memory seeds.


\subsection{Memory Fragments Growth}
\label{fragments}
Although existing system memory refinement approaches can extract key information from retrieved documents, the complex relationships between queries and documents can degrade the quality of both subsequent queries and final answers~\cite{querydocs}.
In particular, the smooth semantic nature of sentence embeddings in vector space causes query representations and lengthy document embeddings to become highly entangled. This entanglement introduces irrelevant information and obscures critical content, thereby reducing the effectiveness of memory refinement. Furthermore, in agentic search, such degraded memories are repeatedly used as inputs for subsequent reasoning steps, compounding errors and leading to a gradual decline in downstream reasoning quality. 

To address this issue, we grow memory fragments that contain the contents of memory seeds introduced in Section~\ref{seeds} based on the retrieved information. 
Instead of comprehending complex semantic relations between the query and retrieved texts, memory seeds guide LLMs to extract diverse and useful summaries aligned with the query goal. 
Concretely, we provide the LLM with the memory seeds $\boldsymbol{s}_i^{(n)}$, retrieval results $\boldsymbol{D}^{(n)}$ of the $n$-th reasoning round and task instructions $\boldsymbol{I}_M$, and prompt it to expand each seed into diverse yet relevant memory fragments for subsequent reasoning. Thus, the generation probability $P_{M}$ of the memory fragment $\boldsymbol{M}^{(n)}$ can be formulated as:
\begin{equation}
\label{memory}
\begin{aligned}
    P_M=\prod_{i=1}^{L_r}\prod_{t=T_{s_{i-1}}^{(n)}}^{T_{s_i}^{(n)}}
P(\boldsymbol{M}_t^{(n)} \vert \boldsymbol{M}_{<t}^{(n)}, \boldsymbol{s}_i^{(n)}, \boldsymbol{I}_M, \boldsymbol{D}^{(n)}),
\end{aligned}
\end{equation}
where $T_{s_i}^{(n)}$ is the position where the $i$-th memory fragment ends, and $T_{s_0}^{(n)}=0$. $\boldsymbol{M}_t^{n}$ denotes the $t$-th generated token of the memory fragments, and $\boldsymbol{M}_{<t}^{(n)}$ is the sequence of the LLM generated before position $t$. The current query $\boldsymbol{q}^{(n)}$ is included in the $\boldsymbol{I}_M$ as a constraint.

Obviously, the $\boldsymbol{s}_i^{(n)}$ serves as a seed that attracts the attention of LLMs to focus on the relevant information to it, which brings about more concise and accurate fragment contents. Subsequently, the generated tokens in Equation \ref{memory} are concatenated to form memory fragments, as formalized below.
\begin{equation}
    \boldsymbol{m}_{i}^{(n)} = \boldsymbol{M}_{T_{s_{i-1}}^{(n)}:T_{s_i}^{(n)}}^{(n)},
\end{equation}
where $\boldsymbol{m}_i^{(n)}$ is the $i$-th grown memory fragment during the $n$-th reasoning round, and $T_{s_0}^{(n)}=0$.

\subsection{Memory Path Retracing}
\label{manifold}
Existing memory management often preserves verbose traces accumulated across multiple reasoning rounds without effective integration, and answering over elongated paths exacerbates memory dilution~\cite{compressing}. We address this by retracing the memory history and reorganizing fragments into a coherent path: MemSearch-o1 selects the most relevant evidence and assembles a semantically smooth memory trajectory, reducing redundancy and sharpening focus for reliable reasoning.

After the LLM completes search and reasoning, we retrace and collect all the memory fragments $\boldsymbol{m}_i$ from every reasoning round within a memory region. However, because search queries may drift semantically over reasoning rounds, the grown fragments can contain irrelevant information. To refine them, we design a contribution function $C(\boldsymbol{m}_i;\boldsymbol{q}_o)$ that measures both their relevance to the original query $\boldsymbol{q}_o$ and their bridge potential to other fragments. Specifically, we define the relevance contribution $C_{Rel}$ to filter out memories that lie far from the ideal reasoning region as follows:
\begin{equation}
    C_{Rel} = \text{Sim}(\text{Emb}(\boldsymbol{m}_i), \text{Emb}(\boldsymbol{q}_o)),
\end{equation}
where $\text{Emb}(\cdot)$ is the sentence transformer that maps input tokens into embedding vectors, and $\text{Sim}(\cdot)$ is the cosine similarity. Memory fragments with high contribution $C_{Rel}$ have more relevance to $\boldsymbol{q}_o$. In addition, we consider the potential of each memory fragment to connect with others through a bridge potential function:
\begin{equation}
    C_{BP} = \frac{
\sum_{j \neq i} \text{sim}(\boldsymbol{m}_i, \boldsymbol{m}_j) \cdot \sigma(U_{Rel} - \tau_s)
}{
\sum_{j \neq i} \sigma(U_{Rel} - \tau_s)
},
\end{equation}
where $U_{Rel}=\max\{\text{Sim}(\boldsymbol{m}_j, \boldsymbol{q}^{(n)})\}_{j=1}^N$ is the upper bound of the relevance between memory fragments and search query, and $N$ is the total number of reasoning rounds. Threshold $\tau_s$ together with the ReLU function $\sigma(\cdot)$ restricts the contribution score of fragments with the lowest relevance to the search query in reasoning. The maximum relevance $\sigma(U_{Rel} - \tau_s)$ serves as the weight on the bridge potential with other fragments, which is calculated by the cosine similarity function. Then we can derive the contribution score of the memory fragment $\boldsymbol{m_i}$ as:
\begin{equation}
    C(\boldsymbol{m}_i;\boldsymbol{q}_o) = \alpha\cdot C_{Rel} + \beta \cdot C_{BP},
\end{equation}
where $\alpha$ and $\beta$ are predefined weights. Then we need to reorganize the fragments that we retrace and select in the memory region:
\begin{equation}
    \mathcal{M}_q = \left\{ \boldsymbol{m}_i \in \mathbb{R}^d \vert  \ C(\boldsymbol{m}_i; \boldsymbol{q}_o) > \tau_r \right\},
\end{equation}
where $\tau_r$ is the threshold that filters the memory fragments with low contribution, i.e., with low semantic relevance to queries and other memory fragments. In order to find a semantically smooth memory path tailored for reasoning, we aim to solve the following optimization problem:
\begin{equation}
\begin{aligned}
    &\mathcal{P}^* = \arg\max_{\mathcal{P}} \sum_{k=1}^{\vert \mathcal{M}_q\vert} C(\boldsymbol{m}_{i_k}; \boldsymbol{q}_o) \cdot \mu(\boldsymbol{m}_{i_k}),\\
    &\mu(\boldsymbol{m}_{i_k}) = \exp\left( -\lambda (1-\text{Sim}(\boldsymbol{m}_{i_k}, \boldsymbol{m}_{i_{k-1}})) \right),
\end{aligned}
\end{equation}
where $\mathcal{P}=(\boldsymbol{m}_{i_1}, \boldsymbol{m}_{i_2}, ..., \boldsymbol{m}_{i_K})$ is the ideal memory path consisting of $K$ maximum fragments, and $\mu(\boldsymbol{m}_{i_k})$ is the penalty function that ensures the smooth semantics between the current fragment and connected fragment. In this work, we utilize the greedy search strategy~\cite{greedy} to find the memory path. Finally, the memory path is directly used for answer generation.

Since the memory fragments are generated based on prior reasoning steps, and we have already organized them through a path that incorporates recall and semantic smoothing, during the generation phase, we no longer need to rely on the original system memory. Instead, we can generate answers directly from the constructed memory path. The detailed deep search algorithm of MemSearch-o1 inference is displayed in Appendix \ref{algorithm}.

\begin{table*}[t]
\centering
\resizebox{\textwidth}{!}{
\begin{tabular}{cc *{9}{c}}
\toprule
 \multirow{2}{*}{Models}& \multirow{2}{*}{Methods} & \multicolumn{4}{c}{\textbf{MultidocQA}} & \multicolumn{4}{c}{\textbf{Singledoc QA}} & \multirow{2}{*}{Avg.} \\
\cmidrule(lr){3-6} \cmidrule(lr){7-10}
 & & HotpotQA & 2WikiMQA & MuSiQue & DuReader & NarrativeQA & Qasper & MultiField-en & MultiField-zh &  \\
\midrule

\multirow{7}{*}{\makecell{\textbf{Qwen 2.5-} \\ \textbf{72B-Instruct}}}
& Direct RAG             & 54.40 & 47.23 & 27.54 & \underline{27.49} & \textbf{17.73} & \underline{35.94} & \underline{45.43} & \underline{52.29} & 38.51 \\
& Search-o1 (RAgent)     & \underline{57.41} & \underline{62.86} & \textbf{44.78} & 18.91 & 14.37 & 22.88 & 37.77 & 49.97 & \underline{38.62} \\
& Search-o1 (Refined) & 52.11 & 35.74 & 37.99 & 15.30 & 16.12 & 17.51 & 33.79 & 44.45 & 31.63 \\
& MemoryBank             & 47.27 & 44.89 & 28.26 & 19.24 & 16.26 & 26.28 & 32.42 & 43.46 & 32.26 \\
& A-Mem                  & 54.24 & 60.08 & 39.95 & 18.74 & 15.18 & 22.09 & 34.59 & 50.79 & 36.96 \\
& Amber                  & 53.78 & 61.16 & 37.61 & 20.57 & 16.65 & 27.19 & 37.98 & 52.26 & 38.40 \\
& \textbf{MemSearch-o1}  & \textbf{59.71}$^{\ast}$ & \textbf{65.95}$^{\ast}$ & \underline{44.11} & \textbf{32.06}$^{\ast}$ & \underline{17.30} & \textbf{36.18}$^{\ast}$ & \textbf{49.05}$^{\ast}$ & \textbf{59.91}$^{\ast}$ & \textbf{45.53}$^{\ast}$ \\
\midrule

\multirow{7}{*}{\textbf{Deepseek V3.1}} 
& Direct RAG             & 53.98 & 43.76 & 26.95 & \underline{18.72} & 21.48 & 32.45 & \underline{43.41} & 48.15 & 36.11 \\
& Search-o1 (RAgent)     & 54.64 & 56.14 & \underline{44.31} & 15.28 & 20.59 & \underline{33.04} & 38.67 & \underline{49.76} & 39.05 \\
& Search-o1 (Refined) & 48.46 & 36.75 & 33.32 & 16.12 & 18.59 & 22.97 & 33.82 & 44.79 & 31.85 \\
& MemoryBank             & 47.51 & 46.65 & 38.05 & 17.57 & 21.83 & 28.60 & 33.34 & 42.33 & 34.49 \\
& A-Mem                  & 49.57 & 41.38 & 25.67 & 17.10 & 20.92 & 27.55 & 28.27 & 47.29 & 32.22 \\
& Amber                  & \underline{55.59} & \underline{58.63} & 41.29 & 18.39 & \underline{22.48} & 32.86 & 39.27 & 49.31 & \underline{39.79} \\
& \textbf{MemSearch-o1}  & \textbf{67.78}$^{\ast}$ & \textbf{68.32}$^{\ast}$ & \textbf{52.01}$^{\ast}$ & \textbf{27.23}$^{\ast}$ & \textbf{23.04}$^{\ast}$ & \textbf{37.94}$^{\ast}$ & \textbf{52.26}$^{\ast}$ & \textbf{56.81}$^{\ast}$ & \textbf{48.17}$^{\ast}$ \\

\bottomrule
\end{tabular}
}
\caption{Overall Performance (\%) Comparison on MultidocQA and Singledoc QA Benchmarks. The best results are bolded and the second best results are underlined. "$^{\ast}$" indicates the statistically significant improvements (i.e.,two-sided t-test with $p<0.05$ ) over the baselines.}
\label{overall}
\vspace{-2mm}
\end{table*}

\section{Experiment}
\subsection{Dataset and Metrics}
To evaluate the effectiveness of MemSearch-o1, we conduct extensive experiments on eight benchmark datasets from LongBench~\cite{longbench}, covering both multi-document and single-document QA. The multi-document tasks include HotpotQA~\cite{hotpotqa}, 2WikiMQA~\cite{2wikimqa}, and MuSiQue~\cite{musique}, DuReader~\cite{dureader}, while the single-document tasks include NarrativeQA~\cite{narrativeqa}, Qasper~\cite{qasper}, and Multi-FieldQA~\cite{longbench}. The long contexts of QA tasks serve as the corpus for external knowledge retrieval, and the retrieved passages are substantially shorter than the average document length in the corpus. We focus on both multi- and single-document QA to comprehensively evaluate the performance of competing methods. Detailed dataset statistics are provided in Appendix~\ref{appendix_dataset}. Following prior work~\cite{longbench}, we report QA-F1 as the primary evaluation metric for all datasets except DuReader, which is evaluated using ROUGE-L. We also evaluate the performance of agentic search baselines on large-corpus search tasks using LongBench v2~\cite{longbenchv2} and LongBook QA~\cite{infinitebench}

\subsection{Baselines}
To verify the effectiveness of our MemSearch-o1 framework, we compare the performance with the following baselines: (1) \textbf{Standard RAG}~\cite{RAG}, (2) Deep search without memory management: \textbf{Search-o1 (RAgent)}~\cite{searcho1} (3) Deep search with memory management: \textbf{Search-o1 (Refined)}~\cite{searcho1}, \textbf{MemoryBank}
~\cite{memorybank}, \textbf{A-Mem}~\cite{amem}, and \textbf{Amber}~\cite{amber}. The detailed information of the baselines and the backbone models can be found in Appendix \ref{baselines}.

\subsection{Implementation Details}
\label{implementation}

For all retrieval processes in the baselines, including standard RAG and deep search systems, we use BGE-M3~\cite{bge} as both the retriever and sentence embedding model. The maximum number of search rounds is fixed at $N=5$, with the top-$k$ documents ($k=3$) retrieved in chunks of 256 tokens. We set filtering thresholds $\tau_s$ and $\tau_r$ to 0.3 to discard memory fragments weakly related to either the original query or reasoning-driven search queries. In the contribution function, weights are set to $\alpha=0.6$ and $\beta=0.4$, while memory reorganization applies a penalty $\lambda=1$ to ensure semantic smoothness and restricts each path to $K=10$ fragments. Following prior work~\cite{searcho1,m+}, we evaluate MemSearch-o1 alongside baselines using two backbone LLMs: Qwen2.5-72B-Instruct~\cite{qwen2.5} and DeepSeek V3.1~\cite{liu2024deepseek}. Full prompt templates are provided in Appendix~\ref{prompt}.

\subsection{Overall Performance}

\noindent\textbf{Experiment on LongBench.}
As shown in Table~\ref{overall}, MemSearch-o1 achieves state-of-the-art results across both multi- and single-document QA benchmarks, outperforming RAG and advanced agentic search methods such as Amber and A-Mem under all two LLM backbones. The gains are especially notable on complex reasoning tasks: with DeepSeek V3.1, MemSearch-o1 improves over the strongest baseline by 21.93\% on HotpotQA, 16.53\% on 2WikiMQA, and 17.38\% on MuSiQue. The results are obtained by averaging over multiple runs. These results show that fine-grained memory growth from seed tokens and retraced memory paths substantially enhance the iterative reasoning quality. To further show that MemSearch-o1 relieves the memory dilution problem and constructs concise memory paths for generation, we measure the number of tokens in the system memory. Detailed analysis can be found in Appendix~\ref{token}.



Deep search with memory management generally surpasses typical RAG on multi-document QA, as iterative retrieval supports multi-hop reasoning. Yet, existing memory strategies remain limited. Summarization-based methods such as Search-o1, MemoryBank, and A-Mem often lose key evidence, while Amber mitigates memory dilution through chunk- and sentence-level refinement but still struggles to capture deeper logical connections.



On tasks with localized evidence (e.g., Qasper, MultiFieldQA-en), standard RAG can be competitive, since these scenarios require less iterative retrieval. Many deep search methods over-search here, leading to instability. In contrast, MemSearch-o1 avoids this issue by expanding memory from seed tokens, ensuring sufficient coverage while maintaining concise and precise reasoning paths. Together, these results establish MemSearch-o1 as a robust framework for deep search reasoning.

To further show the superiority of MemSearch-o1 over agentic search systems equipped with different memory management strategies, we compare it against Search-o1 (with its original memory refinement), MemoryBank~\cite{memorybank}, Zep~\cite{zep}, and MIRIX~\cite{mirix}. The results and detailed analysis are provided in Appendix \ref{memory_management}.

\noindent\textbf{Experiment on LongBench v2 and LongBookQA.}
To further validate the effectiveness of MemSearch-o1, we also conduct experiments on LongBookQA (en\&zh) from InfiniteBench~\cite{infinitebench} and Multi\&Single-Document QA in various knowledge domains from LongBench v2~\cite{longbenchv2}. Specifically, LongBench v2 has the corpus with 15k$\sim$129k tokens, and includes scientific QA tasks covering domains such as academia, finance, government reports, and legal texts. LongBookQA en and zh have a corpus with 192k and 2.068M tokens respectively, and consist of long-form novels and narrative stories. By incorporating these datasets, we demonstrate that MemSearch-o1's capabilities can be successfully extended to a broader range of domains. We use the Qwen2.5-72B-Instruct as the backbone, and employ accuracy to evaluate the multiple-choice questions in Longbench v2, and employ F1 to evaluate LongBook QA. The experimental results of LongBench v2 are shown in Table \ref{tab:multi_domain_results}.

\begin{table*}[t]
    \centering
    \resizebox{0.9\linewidth}{!}{%
    \begin{tabular}{lcccccc}
        \toprule
        \textbf{Methods} & \textbf{Multinews} & \textbf{Academic} & \textbf{Legal} & \textbf{Financial} & \textbf{Governmental} & \textbf{Avg.} \\
        \midrule
        Direct RAG & 30.43 & 33.33 & 24.24 & 29.73 & 26.83 & 28.91 \\
        Search-o1 (Refined) & 30.43 & 27.27 & 32.98 & 32.43 & 31.71 & 30.96 \\
        A-Mem & \underline{39.13} & \textbf{40.43} & \textbf{42.42} & \underline{40.54} & \underline{34.15} & \underline{39.33} \\
        Amber & \underline{39.13} & 39.36 & 36.36 & 35.14 & 29.27 & 35.85 \\
        MemSearch-o1 & \textbf{43.48}$^{\ast}$ & \textbf{40.43} & \textbf{42.42} & \textbf{48.65}$^{\ast}$ & \textbf{36.59}$^{\ast}$ & \textbf{42.31}$^{\ast}$ \\
        \bottomrule
    \end{tabular}%
    }
    \caption{Performance Comparison in LongBench v2. Metrics are Accuracy (\%). The best results are bolded and the second best results are underlined. "$^{\ast}$" indicates the statistically significant improvements (i.e.,two-sided t-test with $p<0.05$ ) over the baselines.}
    \label{tab:multi_domain_results}
    \vspace{-3mm}
\end{table*}

\begin{table}[t]
    \centering
    \resizebox{0.73\linewidth}{!}{
    \begin{tabular}{lcc}
        \toprule
        \textbf{Method} & \textbf{F1} & \textbf{EM} \\
        \midrule
        Direct RAG & 23.13 & 15.67 \\
        Search-o1 (Refined) & 16.91 & 11.68 \\
        A-Mem & 18.64 & 11.40 \\
        Amber & 19.58 & 13.11 \\
        MemSearch-o1 & \textbf{25.04}$^{\ast}$ & \textbf{17.66}$^{\ast}$ \\
        \bottomrule
    \end{tabular}}
    \caption{Results on LongBookQA-en Dataset. The best results are bolded and "$^{\ast}$" indicates the statistically significant improvements (i.e.,two-sided t-test with $p<0.05$ ) over the baselines.}
    \label{tab:longbookqa_en}
\end{table}

\begin{table}[t]
    \centering
    \resizebox{0.73\linewidth}{!}{
    \begin{tabular}{lcc}
        \toprule
        \textbf{Method} & \textbf{F1} & \textbf{EM} \\
        \midrule
        Direct RAG & 37.28 & 30.16 \\
        Search-o1 (Refined) & 22.16 & 18.52 \\
        A-Mem & 28.64 & 23.81 \\
        Amber & 27.54 & 21.69 \\
        MemSearch-o1 & \textbf{39.44}$^{\ast}$ & \textbf{33.86}$^{\ast}$ \\
        \bottomrule
    \end{tabular}}
    \caption{Results on LongBookQA-zh Dataset. The best results are bolded and "$^{\ast}$" indicates the statistically significant improvements (i.e.,two-sided t-test with $p<0.05$ ) over the baselines.}
    \label{tab:longbookqa_zh}
    \vspace{-3mm}
\end{table}

As shown in Table \ref{tab:longbookqa_en} and \ref{tab:longbookqa_zh}, MemSearch-o1 maintains strong deep search performance on extremely large corpora, effectively organizing memory across multiple retrieval rounds to support high-quality reasoning. Additionally, MemSearch-o1 can effectively find and organize the knowledge in the large corpus without training. In contrast, other strong agentic search baselines exhibit unstable performance compared with naive RAG, as they may suffer from oversearch or the misleading search in the large corpus. These observations demonstrate that MemSearch-o1 is highly adaptable to diverse data types, including both scientific QA datasets and narrative/story datasets.

\subsection{Scaling Law of MemSearch-o1}
\begin{figure}[t]
    \centering
    \includegraphics[width=\linewidth]{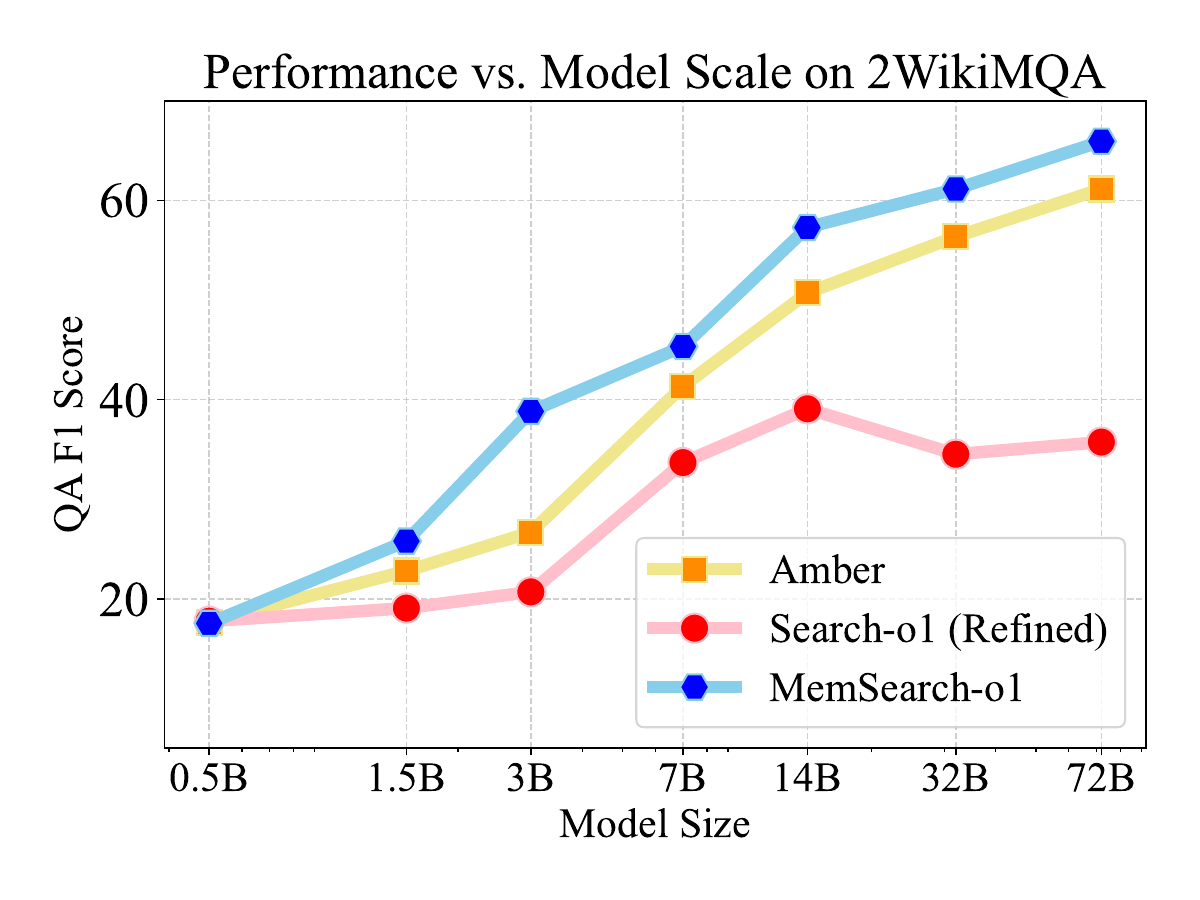}
    \caption{Model Size Scaling on MemSearch-o1}
    \label{scaling}
    \vspace{-3mm}
\end{figure}

In this experiment, we employ the Qwen2.5 instruct model series from 0.5B to 72B to perform the model size scaling and evaluate the performance on the 2WikiMQA dataset. The corresponding results of MemSearch-o1 are shown in Figure \ref{scaling}.


Smaller models struggle to accurately follow search instructions and thus fail to perform deep reasoning without additional training. As model size increases, however, reasoning capabilities are activated more rapidly. Around the 3B scale, our method effectively unlocks deep search-based reasoning, and larger models achieve even higher accuracy, confirming the strong compatibility of our approach with reasoning tasks. In contrast, Amber shows delayed activation, with noticeable improvements only beyond the 7B scale, and still lags behind MemSearch-o1 in overall performance. Search-o1 (Refined), meanwhile, exhibits unstable behavior: after an initial boost, its performance fluctuates on larger models and fails to scale consistently. This instability reflects its reliance on complex semantic associations between queries and retrieved documents, which often leads to information loss and constrains its reasoning capacity.

\subsection{Top-$k$ Scalability for Retrieval}

\begin{figure}[t]
    \centering
    \captionsetup[subfigure]{skip=1pt} 
    
    \begin{subfigure}[b]{\linewidth}
        \centering
        \includegraphics[width=\linewidth]{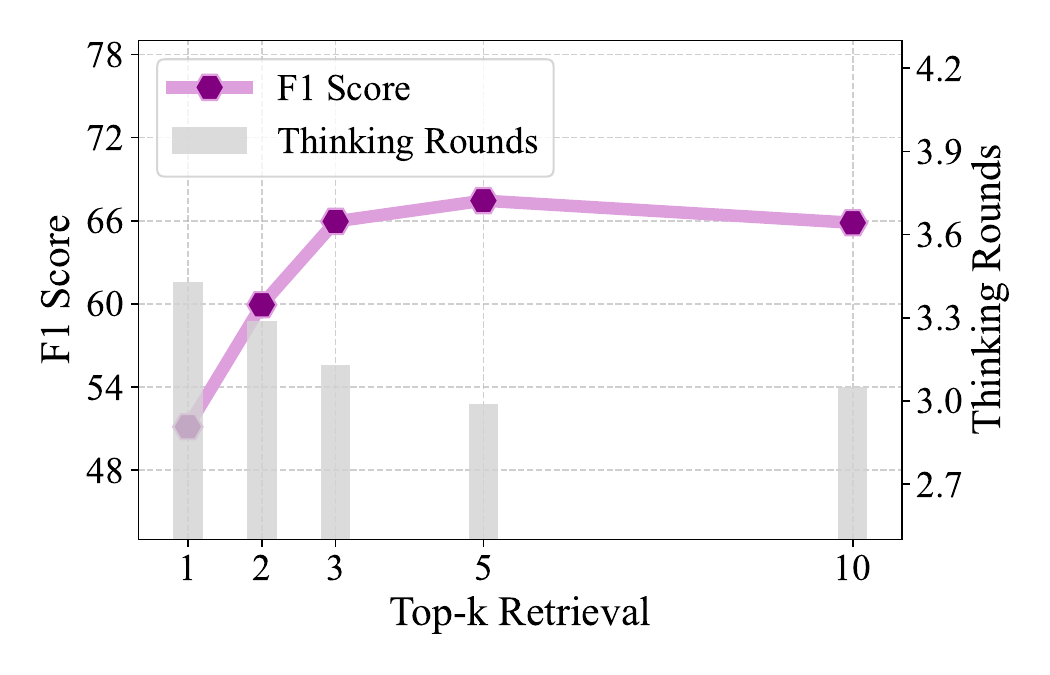}
        \caption{2WikiMQA}
        \label{fig:topk1}
    \end{subfigure}
    
    \vspace{1pt} 
    
    \begin{subfigure}[b]{\linewidth}
        \centering
        \includegraphics[width=\linewidth]{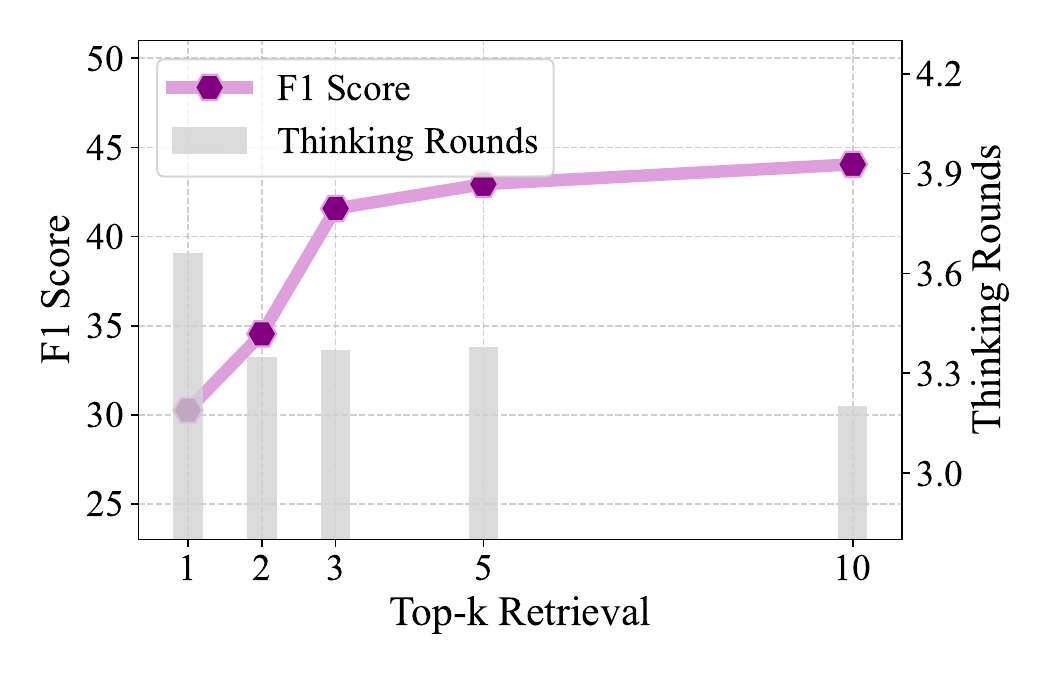}
        \caption{MuSiQue}
        \label{fig:topk2}
    \end{subfigure}
    
    \caption{Top-$k$ Scalability for MemSearch-o1.}
    \label{topk}
    \vspace{-2mm}
\end{figure}
In this subsection, we vary the number of top-$k$ documents retrieved during the agentic search process to examine how retrieval volume influences both the performance of MemSearch-o1 and the average number of reasoning steps. Experiments are conducted with Qwen2.5-72B-Instruct on the 2WikiMQA and MuSiQue datasets, and results are shown in Figure~\ref{topk}.



Two distinct patterns emerge. As illustrated in Figure~\ref{topk}(a), increasing $k$ initially enriches the memory with more relevant information, reducing the number of reasoning steps required. However, excessively large $k$ introduces redundancy, diluting useful context and ultimately reducing accuracy while increasing the number of search rounds. By contrast, in Figure~\ref{topk}(b), performance continues to improve as $k$ grows. For datasets like MuSiQue, where relevant evidence is dispersed across many chunks, larger retrieval sets provide greater information gains for memory construction. Therefore, even though lengthy documents contain abundant information, MemSearch-o1, by employing memory-seed-guided memory growth and focusing on information extraction at the token level, can still distill more effective content from extended documents, thereby reducing the required number of reasoning steps accordingly. Detailed values and additional analysis are provided in Appendix~\ref{scalability}.

\subsection{Ablation Study}

We conduct the ablation study and remove the memory management strategy step by step. The results are shown in Table \ref{tab:ablation}. Specifically, we remove the memory path retracing and reorganization (\emph{w/o memory retracing}) first, to verify the effectiveness of deep refinement in Memsearch-o1. Then, based on this, we remove the memory seeds and fragments growth (\emph{w/o memory}), to show the efficacy of memory management.

The ablation results highlight the importance of memory management in MemSearch-o1. When memory is completely removed, the LLM is exposed to redundant document content, and the complex semantic relations between queries and retrieved texts lead to confused reasoning, causing a noticeable performance drop. Removing the memory path retracing also degrades performance, as unorganized memory fragments accumulate irrelevant information with increasing semantic distance from the original query. These findings confirm that both memory growth and retracing are essential for maintaining reasoning accuracy.
\begin{table}[t]
\centering
\resizebox{\linewidth}{!}{
\begin{tabular}{l c c c}
\toprule
\textbf{Dataset} & \textbf{w/o memory} & \textbf{w/o retracing} & \textbf{Complete} \\
\midrule
hotpotqa           & 54.64 & 64.27 & \textbf{67.78} \\
2wikimqa           & 56.41 & 63.91 & \textbf{68.32} \\
musique            & 44.31 & 47.69 & \textbf{52.01} \\
dureader           & 15.28 & 25.05 & \textbf{27.23} \\
narrativeqa        & 20.59 & 22.50 & \textbf{23.04} \\
qasper             & 33.04 & 35.30 & \textbf{37.94} \\
multifieldqa-en   & 38.67 & 46.33 & \textbf{52.26} \\
multifieldqa-zh   & 49.76 & 53.73 & \textbf{56.81} \\
\bottomrule
\end{tabular}}
\caption{Ablation Study of MemSearch-o1}
\label{tab:ablation}
\vspace{-4mm}
\end{table}

\subsection{Memory Path Vizualization}
\begin{figure}[t]
    \centering
    \begin{subfigure}[b]{\linewidth}
        \centering
        \includegraphics[width=\linewidth]{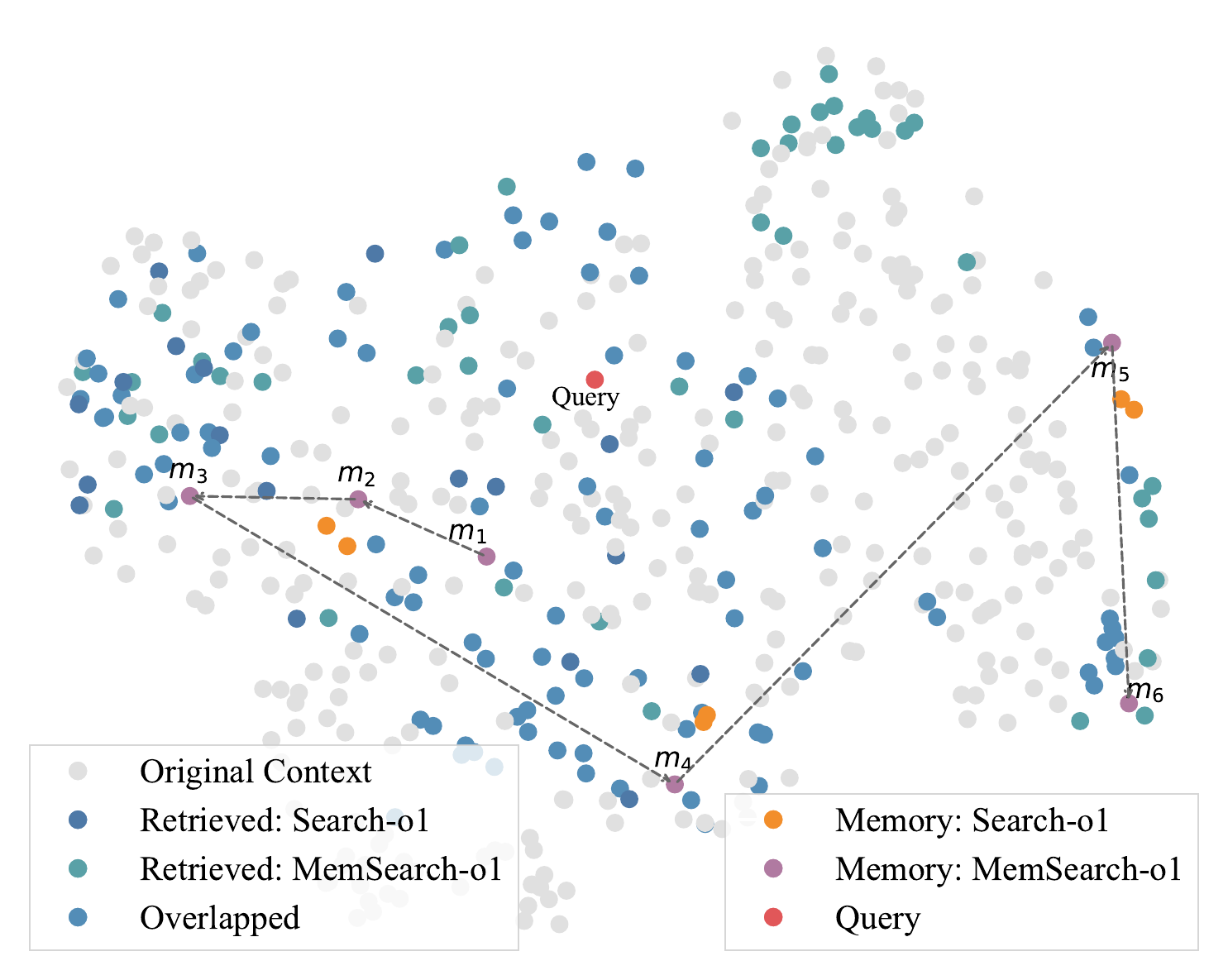}
        \caption{UMAP: Memory Path under Divergent Retrieval Regions}
        \label{fig:umap_top}
    \end{subfigure}
    
    \vspace{2pt} 
    
    \begin{subfigure}[b]{\linewidth}
        \centering
        \includegraphics[width=\linewidth]{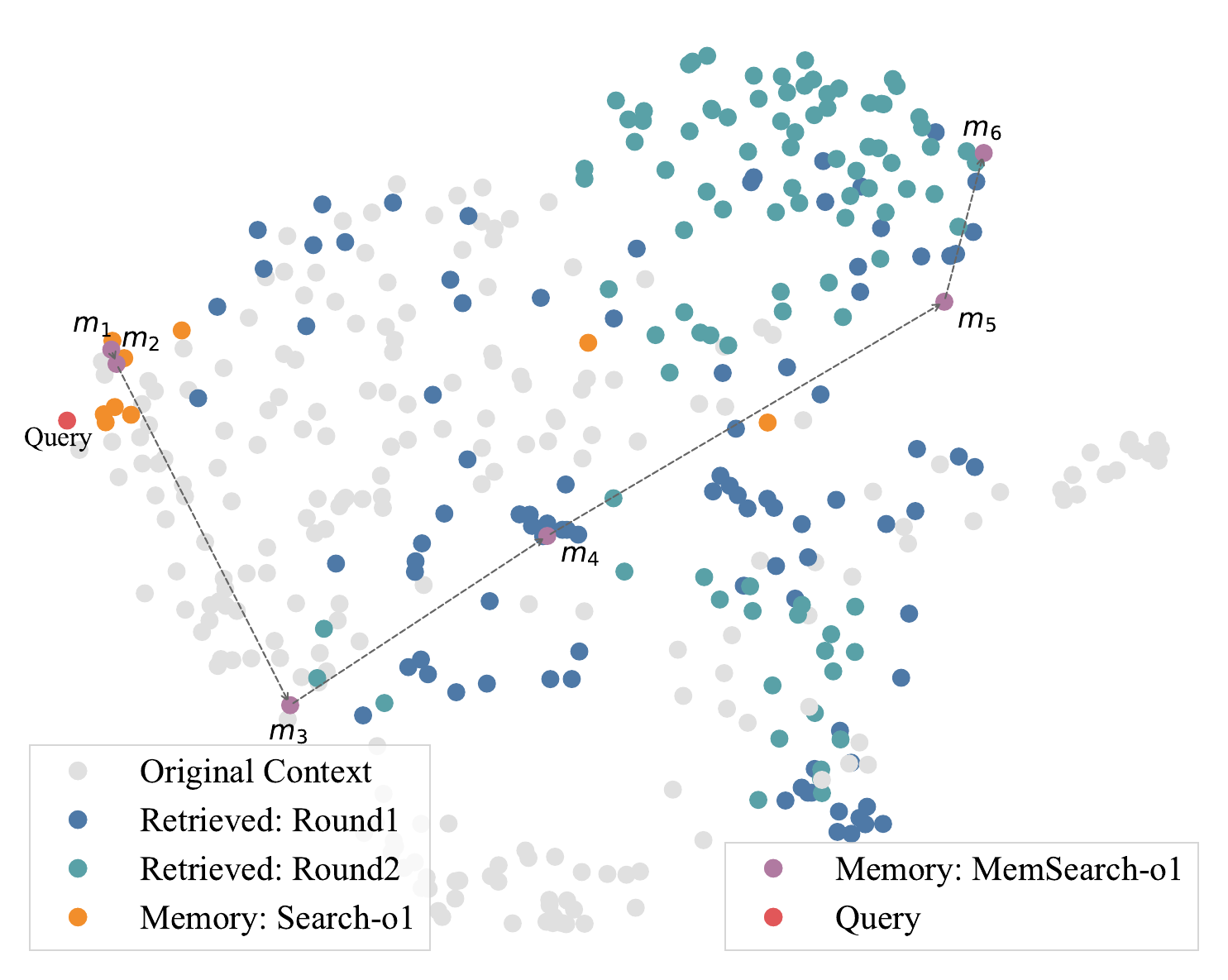}
        \caption{UMAP: Memory Path under Identical Retrieval Regions}
        \label{fig:umap_bottom}
    \end{subfigure}
    
    \caption{Memory Paths for Different Retrieval Cases.}
    \label{umap}
    \vspace{-2mm}
\end{figure}

In this subsection, we present case studies to illustrate how MemSearch-o1 reasons along memory paths, providing an intuitive demonstration of its effectiveness. We compare our method with Search-o1 (Refined) using the Qwen2.5-72B-Instruct backbone. Memory fragments from MemSearch-o1 and analysis sentences from Search-o1, together with retrieved document sentences, are projected into a low-dimensional space using UMAP~\cite{umap}. Figure~\ref{umap} shows two scenarios: (a) when both retrieval queries and retrieved regions differ, and (b) when they are identical.


As shown in Figure~\ref{umap}(a), memory growth allows MemSearch-o1 to generate queries that are more goal-directed and logically coherent, retrieving regions closer to the ground-truth answers. The richer semantics of memory fragments also enable broader exploration within the memory space, improving answer accuracy. In contrast, Figure~\ref{umap}(b) shows that even with identical retrieved regions, MemSearch-o1 outperforms Search-o1 (Refined). While Search-o1 often over-refines and restricts exploration to local information, MemSearch-o1 expands memory into a wider region, supporting more effective reasoning and retrieval.
These visualizations clearly demonstrate how memory growth in MemSearch-o1 guides reasoning and retrieval in a more expansive, coherent, and accurate manner.

\section{Related Work}
\noindent \textbf{Retrieval-Augmented Generation}
RAG~\cite{RAG, li2025towards} has been a widely adopted method for enhancing LLMs with external knowledge~\cite{raido, PBR, xu2025align, gu2026nanoknow}. Single-step retrieval methods such as GraphRAG~\cite{graphrag}, HippoRAG~\cite{hipporag}, and HiRAG~\cite{HiRAG} leverage knowledge graphs or hierarchical structures to better exploit retrieved contexts. However, their limited retrieval scope often fails on complex multi-hop reasoning tasks~\cite{amber}. To address this, deep search has been proposed as an agentic RAG framework that integrates reasoning, planning, and retrieval~\cite{searcho1, atomsearcher, reasonrag, zhang2026search, chen2025browsecomp}, enabling iterative exploration. Yet, existing methods frequently generate redundant documents and noisy reasoning trajectories, leading to memory dilution and reduced effectiveness~\cite{lost_middle}. In this work, we tackle this issue by introducing memory growth and retracing to construct compact, explicit memories that support high-quality reasoning. Unlike R1-style methods~\cite{searchr1, Graphr1, dynasearcher, autorefine} based on smaller LLMs, our focus is on fully activating the reasoning potential of large-scale models.

\noindent \textbf{Memory Management}
To mitigate memory dilution, prior work has explored a range of memory management strategies for LLMs~\cite{deng2026enhancing, wenmemory, evoking, derongmem}. Search-o1~\cite{searcho1} refines retrieved texts to retain the most useful information, while MemoryBank~\cite{memorybank} summarizes event histories and applies selective forgetting. Mem0~\cite{mem0} combines multi-step retrieval with summarization to enrich semantic coverage, and Amber~\cite{amber} filters relevant content at both chunk and sentence levels during memory updates. Despite these advances, such methods largely rely on summarization or retrieval of query-relevant texts~\cite{mem1}, leaving LLMs unable to fully capture the complex semantic associations between evolving queries and retrieved contexts. This significantly constrains their deep search potential. In this work, we address this limitation by constructing concise yet semantically rich system memories, thereby unlocking more scalable deep search capabilities for LLMs.
\section{Conclusion}
In this paper, we propose MemSearch-o1, a novel deep search system that grows memory fragments from seeds and seeks a path, enhancing the deep search reasoning process, which significantly relieves the problem of memory dilution. This enables the deep search system to effectively extract concise information with rich semantics, and explore the memory path where the LLMs can search for and generate better solutions. Through extensive experiments on eight benchmark datasets, we demonstrate the effectiveness of MemSearch-o1 in enhancing the agentic search capability of LLMS, surpassing all the baselines, including RAG and agentic search frameworks with memory management. Our MemSearch-o1 aligned with reasoning based on memory growth establishes a new foundation for memory-aware agentic intelligence.
\section*{Limitations}
While MemSearch-o1 effectively grows memory fragments and retraces memory paths for reasoning, it still depends on the comprehension ability of LRMs for memory seed preparation and growth. Our method demonstrates substantial improvements with large-parameter LLMs, but the gains on smaller models (fewer than 3B parameters) remain limited, as these models lack strong reasoning and information extraction capabilities and often fail to follow search instructions reliably. Enhancing the performance of small-scale models thus remains an important avenue for future work. 


\section*{Acknowledgements}
This research was partially supported by National Natural Science Foundation of China (No.62502404), Hong Kong Research Grants Council (Research Impact Fund No.R1015-23, Collaborative Research Fund No.C1043-24GF, General Research Fund No. 11218325), Institute of Digital Medicine of City University of Hong Kong (No.9229503), and Huawei (Huawei Innovation Research Program).

\bibliography{custom}

\appendix

\section{Appendix}

\subsection{Algorithm}
\label{algorithm}
The detailed inference procedure of MemSearch-o1 is shown in Algorithm \ref{alg:algorithm}. For MemSearch-o1 inference, we should prepare the model with reasoning ability $\mathcal{R}$, search function \texttt{Search}, toolkit for memory seed preparation $\mathcal{T}$. We input the original query, and instructions for reasoning, memory growth and answers to the LLM to execute inference. Before the inference starts, we initialize the maximum search round as $N$, and the current query $\boldsymbol{q}$ as the original $\boldsymbol{q}_o$ for first-step analysis. While the current search round is less than $N$, and the search indicator is True, the LLM should generate a sequence of analysis using current search query. In line2, if the reasoning sequence of the current round $\boldsymbol{R}^{(n)}$ ends with $\langle\text{|end\_search\_query|}\rangle)$, this means the model $\mathcal{R}$ should execute search to seek for more information. Then, the raised query $\boldsymbol{q}^{(n)}$ is extracted from $\boldsymbol{R}^{(n)}$, and is utilized to retrieve relevant contexts in line 5. In line 6, we use the toolkit to split the query and prepare for memory seed tokens which is stored in the set $\mathcal{S}^{(n)}$. In line 7, we grow memory fragments using the LLM based on the intent of current query $\boldsymbol{q}^{(n)}$.
If the reasoning sequence $\boldsymbol{R}^{(n)}$ ends with EOS, the model regards the retrieved information as sufficient, and finishes the reasoning process. Finally, we should consolidate the memory fragments and calculate the contribution function to find a memory region in line 11 and 12. In line 13, we use greedy search to deeply refine and reorganize the memory path for reasoning, as the some memory fragments might have the semantics far away from that of the original query. Finally in line 14, we use the model $\mathcal{R}$ to generate the answer $a$. 
\begin{algorithm*}[t]
    \caption{MemSearch-o1 Inference}
    \begin{flushleft}
        \textbf{Require:} LLM for reasoning $\mathcal{R}$, search function \texttt{Search}, toolkit for memory seed preparation $\mathcal{T}$.\\
        \textbf{Input:} Original query $\boldsymbol{q}_o$, task instruction $\boldsymbol{I}_R$, answer instruction $\boldsymbol{I}_a$, memory growth instruction $\boldsymbol{I}_M$\\
        Initialize the search indicator $\mathcal{F}=\text{True}$, Corpus $\mathcal{C}$ for retrieval and maximum round of search $N$.\\
        Initialize the memory seed set $\mathcal{S}$ and memory fragment set $\mathcal{M}$ as empty, and set search round $n=1$.\\
        Initialize the current query $\boldsymbol{q}$ for analysis and retrieval as $\boldsymbol{q}_o$, and the reasoning text $\boldsymbol{R}^{(0)}$ as empty.
    \end{flushleft}
    \begin{algorithmic}[1]
        \While{$n\leq N$ and $\mathcal{F} = \text{True}$}
        \State Generate reasoning and derive an analysis sequence. $\boldsymbol{R}^{(n)} \leftarrow \mathcal{R}(\boldsymbol{I}_R, \boldsymbol{q}, \boldsymbol{R}^{(n-1)})$
        \If{$\boldsymbol{R}^{(n)}$ ends with $\langle\text{|end\_search\_query|}\rangle$}
        \State Extract search query: $\boldsymbol{q}^{(n)} \leftarrow \text{Extract}(\boldsymbol{R}^{(n)}, \langle\text{|begin\_search\_query|}\rangle, \langle\text{|end\_search\_query|}\rangle)$\;
        \State Retrieve documents: $\boldsymbol{D} \leftarrow \texttt{Search}(\boldsymbol{q}^{(n)})$
        \State Prepare memory seed tokens $\mathcal{S}^{(n)} = \big\{ \boldsymbol{s}_1^{(n)}, \boldsymbol{s}_2^{(n)}, \dots, \boldsymbol{s}_{L_r}^{(n)} \big\} \;\xleftarrow\;  \mathcal{T}(\boldsymbol{q}^{(n)})$
        \State Grow memory fragments $\mathcal{M}^{(n)}=\{\boldsymbol{m}_1^{(n)}, \boldsymbol{m}_2^{(n)}, \dots, \boldsymbol{m}_{\vert \mathcal{M}^{(n)}\vert}^{(n)}\}\leftarrow \mathcal{R}(\mathcal{S}^{(n)}, \boldsymbol{I}_M, \boldsymbol{q}^{(n)})$.
        \State $n = n+1$
        \ElsIf{$\boldsymbol{R}^{(n)}$ ends with EOS}
        \State Finish the Reasoning Step. $\mathcal{F}=\text{False}$
        \State Memory consolidation: $\mathcal{M}=\{\boldsymbol{m}_1, \boldsymbol{m}_2, \dots, \boldsymbol{m}_{\vert \mathcal{M}\vert}\}\leftarrow \bigcup_{n=1}^N \mathcal{M}^{(n)}$
        \State Calculate the contribution to find memory region: $\mathcal{M}_q=\lbrace \boldsymbol{m_i \in \mathbb{R}^d \vert }C(\boldsymbol{m}_i;\boldsymbol{q}_o)>\tau_r\rbrace$.
        \State Find an optimal memory path $\mathcal{P}^{\ast}=(\boldsymbol{m}_{j_1}, \boldsymbol{m}_{j_2}, ..., \boldsymbol{m}_{j_K})$ using greedy search.
        \State Generate the final response: $\boldsymbol{a} \leftarrow \mathcal{R}(I_a, \boldsymbol{q}_o, \mathcal{P}^{\ast})$
        \EndIf
        \EndWhile
    \end{algorithmic}
    \begin{flushleft}
        \textbf{Output:} High quality reasoning chain $\boldsymbol{R}$ and answer $\boldsymbol{a}$
    \end{flushleft}
    \label{alg:algorithm}
\end{algorithm*}

\subsection{Details for the Datasets}
\label{appendix_dataset}
In our experiments, we utilize LongBench~\cite{longbench} to evaluate the performance of methods, which covers both Chinese and English tasks, offering a more complete assessment of how well models handle questions and comprehend information in long contexts in different languages. We select MultiDocQA and SingleDocQA tasks for evaluation, and the brief description of adopted domains are listed as below:

\noindent \textbf{HotpotQA}~\cite{hotpotqa}: HotpotQA is a question-answering dataset that includes natural, multi-step questions. It provides clear labels for the supporting facts, helping build question-answering systems that are easier to explain.

\noindent \textbf{2WikiMQA}~\cite{2wikimqa}: 2WikiMQA is a high-quality multi-hop QA dataset designed to evaluate a model’s reasoning and inference capabilities. The dataset incorporates evidences by combining structured and unstructured data, and explicitly provides reasoning paths for multi-hop questions.

\noindent \textbf{MuSiQue}~\cite{musique}: MuSiQue is a challenging multi-hop QA dataset constructed via a bottom-up approach that composes connected single-hop questions, ensuring that each reasoning step depends on information from another.

\noindent \textbf{DuReader}~\cite{dureader}: DuReader is a human-annotated Chinese machine reading comprehension dataset grounded in real-world QA tasks. It emphasizes authenticity by featuring genuine user queries, naturally occurring documents, human-provided answers, and practical application contexts.

\noindent \textbf{NarrativeQA}~\cite{narrativeqa}: NarrativeQA is a reading comprehension dataset designed to evaluate deep, integrative understanding of long-form narratives, such as books and movie scripts. 

\noindent \textbf{Qasper}~\cite{qasper}: Qasper is a QA dataset specifically designed for NLP papers. Questions are authored by NLP practitioners who only read the title and abstract, yet seek information that resides in the full paper; answers and supporting evidence are then provided by other experts. 

\noindent \textbf{MultiFieldQA}~\cite{longbench}: MultiFieldQA is a manually curated QA dataset designed to evaluate long-context comprehension across diverse domains, with versions in English and Chinese.It draws documents from legal texts, government reports, encyclopedias, and academic papers.

The statistics of the above domains for evaluation are shown in Table \ref{datset}. The above statistics show that our evaluation is conducted on a diverse set of datasets, ranging from those with shorter contexts where query-relevant information is easier to retrieve—to those with much longer contexts, posing greater challenges for information retrieval. MultiDocQA tasks require synthesizing evidence across multiple, often disjoint, passages to answer complex, multi-hop questions, while SingleDoc QA focuses on precise comprehension and reasoning within a single, typically longer, document.

\subsection{Baselines}
\label{baselines}
In this subsection, we will introduce the baselines adopted in our experiments in detail. 

\noindent \textbf{Direct RAG}~\cite{RAG}: Retrieve relevant documents as external knowledge from the corpus to enhance the the generation quality of LLMs.

\noindent \textbf{Search-o1 (RAgent)}~\cite{searcho1}: Search-o1 as a pioneering deep search system, enables LLMs to integrate agentic search into o1-like reasoning process, during which allows the search agent to reason within the retrieved documents.

\noindent \textbf{Search-o1 (Refined)}~\cite{searcho1}: This is a deep search system integrating memory refinement mechanism into reasoning. It summarizes the retrieved contexts at each search step to improve the quality and efficiency of reasoning.

\noindent \textbf{MemoryBank}~\cite{memorybank}: A typical memory management method, and is applied to search-o1's memory in our experiment. It summarizes the key events from the retrieved texts, and employs forgetting mechanism properly.

\noindent \textbf{A-Mem}~\cite{amem}: A novel memory management method, and is applied to deep search system in our experiments. It generates key notes for retrieved documents, and evolve the memories using links between the notes.

\noindent \textbf{Amber}~\cite{amber}: A deep search system using a novel system memory management method, which filters the retrieved documents in chunk and sentence levels to optimizes the reasoning process.

\begin{table}[t]
\centering
\resizebox{\linewidth}{!}{
\begin{tabular}{cccccc}
\toprule
\textbf{Task} & \textbf{Metric} & \textbf{Avg. Length} & \textbf{Language} & \textbf{\#Samples} \\
\midrule
HotpotQA & F1 & 9,151 & EN & 200 \\
2WikiMultihopQA & F1 & 4,887 & EN & 200 \\
MuSiQue  & F1 & 11,214 & EN & 200 \\
DuReader & Rouge-L & 15,768 & ZH & 200 \\
MultiFieldQA-en & F1 & 4,559 & EN & 150 \\
MultiFieldQA-zh & F1 & 6,701 & ZH & 200 \\
NarrativeQA & F1 & 18,409 & EN & 200 \\
Qasper& F1 & 3,619 & EN & 200 \\
\bottomrule
\end{tabular}}
\caption{Statistics of Evaluation Datasets}
\label{datset}
\vspace{-5mm}
\end{table}

\begin{table*}[t]
\centering
\resizebox{\textwidth}{!}{
\begin{tabular}{cc *{9}{c}}
\toprule
 \multirow{2}{*}{Models}& \multirow{2}{*}{Methods} & \multicolumn{4}{c}{\textbf{MultidocQA}} & \multicolumn{4}{c}{\textbf{Singledoc QA}} & \multirow{2}{*}{Avg.} \\
\cmidrule(lr){3-6} \cmidrule(lr){7-10}
 & & HotpotQA & 2WikiMQA & MuSiQue & DuReader & NarrativeQA & Qasper & MultiField-en & MultiField-zh &  \\
\midrule

\multirow{5}{*}{\makecell{\textbf{Qwen 2.5-} \\ \textbf{72B-Instruct}}}
& Search-o1 (Refined) & 52.11 & 35.74 & 37.99 & 15.30 & 16.12 & 17.51 & 33.79 & 44.45 & 31.63 \\
& MemoryBank & 47.27 & 44.89 & 28.26 & 19.24 & 16.26 & 26.28 & 32.42 & 43.46 & 32.26 \\
& MIRIX & 55.20 & 62.49 & 39.76 & 15.74 & 17.07 & 19.27 & 35.90 & 48.06 & 36.69 \\
& Zep & 56.68 & 60.85 & 41.28 & 15.11 & 14.79 & 17.69 & 36.54 & 49.70 & 36.58 \\
& \textbf{MemSearch-o1}  & \textbf{59.71}$^{\ast}$ & \textbf{65.95}$^{\ast}$ & \textbf{44.11}$^{\ast}$ & \textbf{32.06}$^{\ast}$ & \textbf{17.30}$^{\ast}$ & \textbf{36.18}$^{\ast}$ & \textbf{49.05}$^{\ast}$ & \textbf{59.91}$^{\ast}$ & \textbf{45.53}$^{\ast}$ \\
\bottomrule
\end{tabular}
}
\caption{Detailed comparison with diverse memory management strategies. The best results are bolded. "$^{\ast}$" indicates the statistically significant improvements (i.e.,two-sided t-test with $p<0.05$ ) over the baselines.}
\label{tab:memory_management}
\vspace{-4mm}
\end{table*}

\subsection{Detailed Comparison with Memory Management Strategies.}
\label{memory_management}
In this section, we will compare the performance of MemSearch-o1 over agentic search systems equipped with different memory management strategies. These strategies include step-wise refinement employed in basic Search-o1 (Refined), MemoryBank~\cite{memorybank}, Zep~\cite{zep}, and MIRIX~\cite{mirix}. MIRIX streamlines memory retrieval and management by using memory agents, whereas Zep places greater emphasis on entity relationships within retrieved texts. The experimental results on the LongBench benchmark are shown in Table \ref{tab:memory_management}. The experimental settings of Zep and MIRIX are identical to the basic settings illustrated in the Section \ref{implementation}. We utilize Qwen2.5-72B-Instruct as the agent model.

The results show that traditional memory management strategies may struggle to grasp complex semantics in the retrieved documents. This is because the refinement- and retrieval-based methods may ignore important information. Additionally, graph-based methods may degrade deep reasoning performance, as inferring over graph-structured knowledge constitutes a challenging multi-hop reasoning task, particularly when the retrieved context contains significant noise. Uniquely, in contrast to these approaches, MemSearch-o1 is a novel method with stronger refinement performance that leverages part-of-speech tagging to grow and trace back memory starting from memory seed tokens derived from the query. This enables LLMs to directly identify and organize crucial memory information within lengthy retrieved documents.

\subsection{Average Tokens for Generation}
\label{token}
\begin{figure}[t]
    \centering
    \begin{subfigure}[b]{0.9\linewidth}
        \centering
        \includegraphics[width=\linewidth]{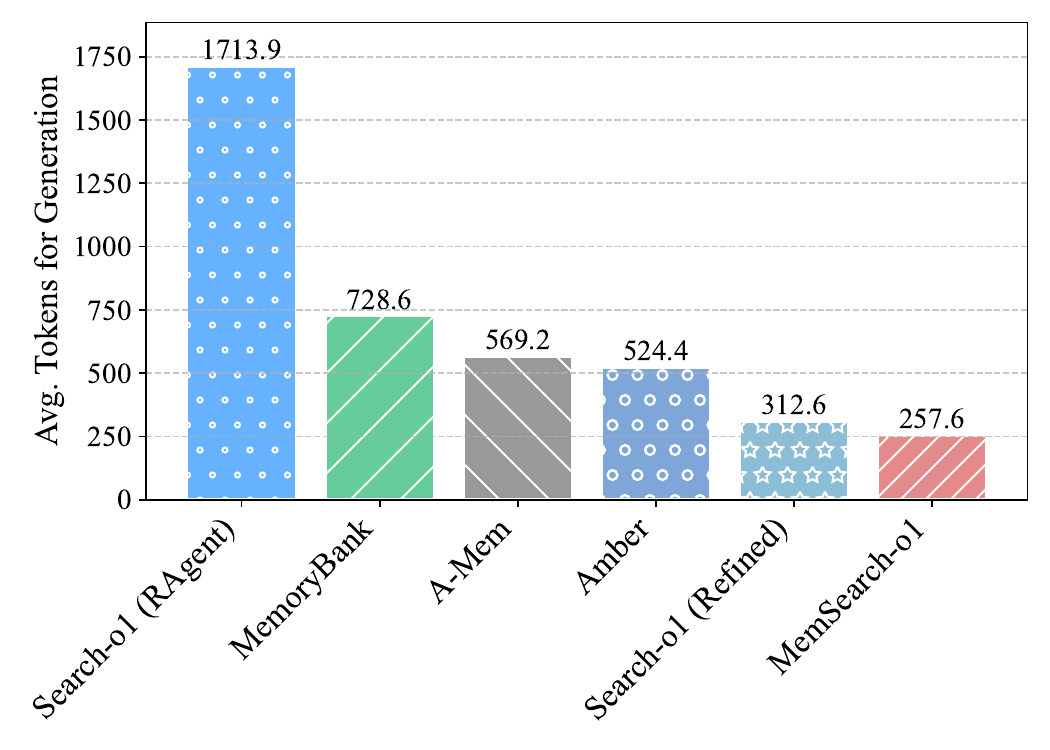}
        \caption{HotpotQA.}
        \label{fig:dilution_top}
    \end{subfigure}
    
    \vspace{2pt} 
    
    \begin{subfigure}[b]{0.9\linewidth}
        \centering
        \includegraphics[width=\linewidth]{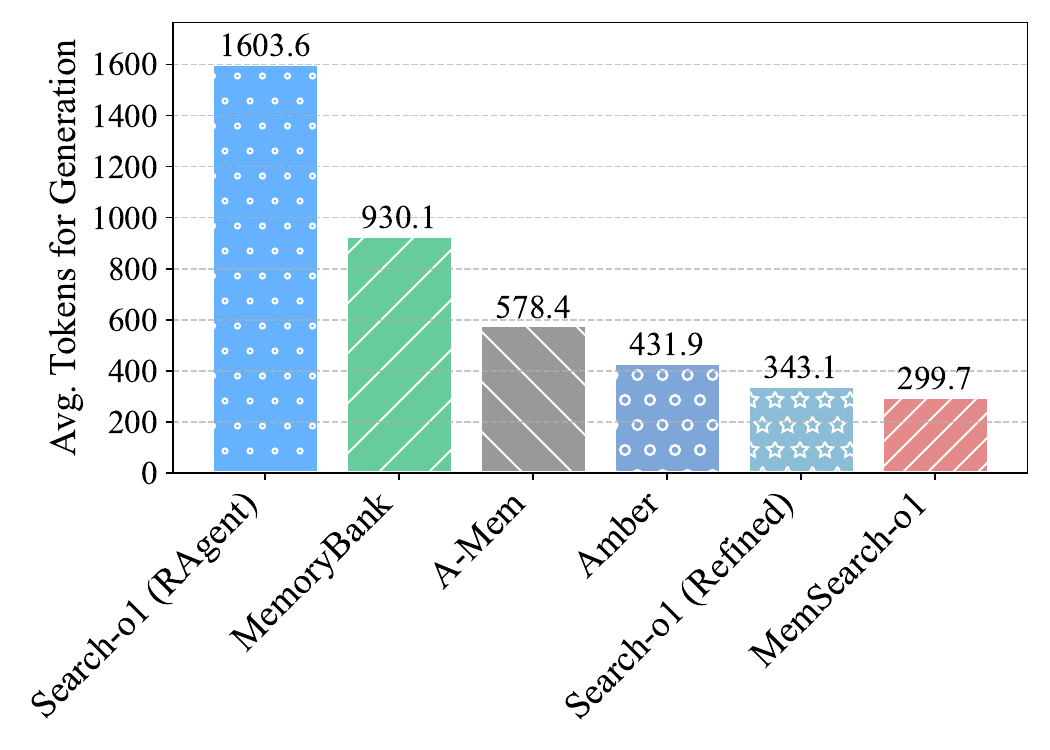}
        \caption{2WikiMQA.}
        \label{fig:dilution_bottom}
    \end{subfigure}
    
    \caption{Average Tokens for Generation.}
    \label{dilution}
\end{figure}

In this subsection, we show the average tokens for generating answers across all baseline methods. The experiments are conducted on HotpotQA and 2WikiMQA, using the DeepSeek-V3.1 backbone. We limit the maximum number of search iterations to three. The results are displayed in Figure~\ref{dilution}. 

The experimental results show that refining retrieved texts without careful processing leads to verbose memory representations, causing severe memory dilution and thereby increasing the difficulty for LLMs to generate accurate responses. This finding is consistent with the instability of Search-o1 (RAgent) observed in Table~\ref{overall}. Directly applying the document refinement strategy from Search-o1, however, tends to over-summarize the content, resulting in significant information loss, as such refinement often overlooks the complex semantics of the queries generated during deep search. Although other memory refinement approaches, such as A-Mem and Amber, achieve relatively better refinement performance, they still extract irrelevant information from documents, ultimately degrading generation quality. Uniquely, our MemSearch-o1 method eliminates the reliance on reasoning history inherent in traditional deep search frameworks. Instead, it grows memory structures from memory seed tokens in a targeted way and then identifies optimal memory paths from the grown memory fragments for response generation. This approach substantially reduces the number of tokens required during generation, improving both efficiency and generation quality while effectively mitigating the memory dilution problem.

We also compare the average total token consumption and the average inference time of each question on the Hotpotqa dataset using Qwen2.5-72B-Instruct as the backbone.

As shown in Table \ref{tab:efficiency_comparison}, MemSearch-o1 achieves lower total token usage across the reasoning pipeline. The inference time of MemSearch-o1 is lower compared with existing approaches, despite its enhanced memory reasoning capability. Moreover, MemSearch-o1 exhibits low time complexity. Unlike traditional deep search methods, which require reading a cumulative system memory with length $D$ in $N$ reasoning steps, leading to $O(N^2D)$ complexity, our approach only extracts memory information from the documents retrieved in the current turn. This design results in complexity $O(ND)$, which means MemSearch-o1 only needs to read the lengthy retrieved documents $N$ times. Consequently, as the number of search iterations increases, the efficiency advantage of MemSearch-o1 becomes more obvious, making it particularly well-suited for multi-hop or iterative reasoning scenarios that demand both accuracy and scalability.

\subsection{Detailed Values of Top-$k$ Scalability}
\label{scalability}
In this subsection, we supplement the detailed values of the top-$k$ scalability, and analyze the mechanism behind the observed results.

As shown in Table \ref{tab:topk_analysis}, as $k$ increases, a richer set of information sources will be retrieved for memory construction. Consequently, during the memory seed expansion process, the growing memory fragments benefit from more contextual input, resulting in semantically richer representations. As a result, the deep search performance improves rapidly in the initial phase of increasing $k$. Moreover, the higher-quality retrieved information boosts the model’s confidence, leading to a reduction in the number of search iterations. This phenomenon is especially obvious on the results of MuSiQue dataset. However, as the retrieved information becomes extremely large, redundant texts will be fed into the LLMs. Since the memory fragments growth are based on both memory seeds and the retrieved texts, too lengthy texts may also lead to the context dilution, and guide the memory growth to irrelevant information region.

\begin{table}[t]
\centering
\resizebox{\linewidth}{!}{
\begin{tabular}{c *{2}{c} *{2}{c}}
\toprule
\multirow{2}{*}{\textbf{top-$k$}} & 
\multicolumn{2}{c}{\textbf{2WikiMQA}} & 
\multicolumn{2}{c}{\textbf{MuSiQue}} \\
\cmidrule(lr){2-3} \cmidrule(lr){4-5}
& \textbf{F1} & \textbf{Thinking rounds} & \textbf{F1} & \textbf{Thinking rounds} \\
\midrule
1 & 51.14 & 3.43 & 30.26 & 3.66 \\
2 & 59.94 & 3.29 & 33.54 & 3.35 \\
3 & 65.95 & 3.13 & 41.57 & 3.37 \\
5 & 67.45 & 2.99 & 42.94 & 3.38 \\
10 & 65.85 & 3.05 & 44.05 & 3.20 \\
\bottomrule
\end{tabular}}
\caption{Performance and Reasoning Cost vs. Top-$k$ Retrieval on 2WikiMQA and MuSiQue}
\label{tab:topk_analysis}
\end{table}

\begin{table}[t]
    \centering
    \resizebox{\columnwidth}{!}{%
    \begin{tabular}{lcc}
        \toprule
        \textbf{Method} & \textbf{Inference Time (s)} & \textbf{Total Tokens} \\
        \midrule
        Search-o1 (Refined) & 22.83 & 2677.6 \\
        A-Mem & 27.07 & 3185.3 \\
        MIRIX & 32.34 & 3956.3 \\
        Zep & 28.33 & 3580.2 \\
        Amber & 20.29 & 2507.3 \\
        MemSearch-o1 & \textbf{18.17} & \textbf{2171.9} \\
        \bottomrule
    \end{tabular}%
    }
    \caption{Comparison of average inference time and total token consumption on HotpotQA.}
    \label{tab:efficiency_comparison}
    \vspace{-4mm}
\end{table}


\subsection{Prompts}
\label{prompt}
The instructions for agentic search follow the previous work~\cite{searcho1, m+}. We take the HotpotQA as an example, and the instructions for deep search are shown in Table \ref{tab:assistant_prompt}. For the memory fragments growth, we organize the prompt template in Table \ref{tab:extraction_prompt}. The prompts for answer generation is shown in Table \ref{tab:qa_prompt}. In addition, for the LongBench v2 benchmark, the specific prompt used to generate final answers for multiple-choice questions is explicitly shown in Table \ref{tab:longbench_v2_prompt}. For the LongBook QA benchmark, we utilize the same experimental settings and prompts as those used for LongBench in our paper.

Regarding prompt design, we follow the experimental setup of prior work, such as Search-o1, and the prompt of our approach only adds a memory fragment extraction step. This memory fragment growth imposes lower difficulty on large language models and does not require highly complex or fine-tuned prompt engineering. Moreover, the extraction of memory fragments is guided by part-of-speech-based memory seed tokens and task-specific targets, enabling the model to reliably extract relevant and essential information across a wide variety of tasks. This design ensures that our method remains effective and generalizable even in more complex and diverse task settings.

Each path constructed by these prompts is an ordered sequence of memory fragments, where the score of a candidate fragment depends on the previously selected one. MemSearch-o1 ensures high-quality construction by two mechanisms: (1) Fragments are filtered via a contribution-based scoring function. (2) The path length is not lengthy, lowering the risk of low-performance of the memory path with high-quality fragments. In short, MemSearch-o1 organizes compact, coherent fragments to effectively structure contextual knowledge and support high-quality reasoning and generation in LLMs. 

\begin{table}[t]
\centering
\resizebox{\linewidth}{!}{
\begin{tabular}{ccccccc}
\toprule
$\alpha$ & $\beta$ & $K$ & $\tau$ & $\lambda$ & \textbf{MuSiQue} & \textbf{MultiField-zh} \\
\midrule
0.8 & 0.2 & 10 & 0.3 & 1 & 43.42 & 58.73 \\
0.6 & 0.4 & 15 & 0.3 & 1 & 43.57 & 59.54 \\
0.6 & 0.4 & 10 & 0.1 & 1 & 43.84 & 59.12 \\
0.6 & 0.4 & 10 & 0.3 & 2 & 43.23 & 58.55 \\
\bottomrule
\end{tabular}}
\caption{Hyperparameter Analysis of $\alpha, \beta, K, \tau$ and $\lambda$ on MuSiQue and MultiField-zh Benchmarks}
\label{tab:hyperparams_ablation}
\vspace{-5mm}
\end{table}

\subsection{Hyperparameter Analysis}
A reasonable hyperparameter setting might not be optimal for all the datasets, but MemSearch-o1 can remain high performance with this setting. In Section \ref{implementation}, we reported that the values of these parameters remain unchanged across all the benchmark datasets. Using this effective setting, MemSearch-o1 exhibits significantly better performance than baselines, indicating it is not dataset-specific.

To further demonstrate that the settings are not dataset-specific, we select some reasonable settings in (1) and use them to test on other datasets. The results are shown in Table \ref{tab:hyperparams_ablation}.

From the results in this table, it can be found that the hyperparameters in reasonable ranges leads to stable performance, and are superior to the baseline methods, which indicates that our hyperparameter settings are not dataset-specific.

\begin{table*}[t]
\centering
\begin{tabular}{|p{0.95\linewidth}|}
\hline
You are a reasoning assistant with the ability to perform searches to help you answer the user's question accurately. When answering, just give the answer and do not output other information. \\
You have special tools: \\
- To perform a search: write <|begin\_search\_query|> your query here <|end\_search\_query|>. \\
Then, the system will search and analyze relevant passages, then provide you with helpful information in the format <|begin\_search\_result|> ...search results... <|end\_search\_result|>. \\
If you think the searched information is not enough, you can continue searching. The maximum number of search attempts is limited to \{MAX\_SEARCH\_LIMIT\}. \\
Once you have all the information you need, stop the search and continue your reasoning. \\
Example: \\
Question: ``Alice David is the voice of Lara Croft in a video game developed by which company?'' \\
Assistant thinking steps: \\
- I need to find out who voices Lara Croft in the video game. \\
- Then, I need to determine which company developed that video game. \\
Assistant: \\
<|begin\_search\_query|>Alice David Lara Croft voice<|end\_search\_query|> \\
(System returns processed information from relevant passages) \\
Assistant thinks: The search results indicate that Alice David is the voice of Lara Croft in a specific video game. Now, I need to find out which company developed that game. \\
Assistant: \\
<|begin\_search\_query|>video game developed by Alice David Lara Croft<|end\_search\_query|> \\
(System returns processed information from relevant passages) \\
Assistant continues reasoning with the new information... \\
Remember: \\
- Use <|begin\_search\_query|> to request a search and end with <|end\_search\_query|>. \\
- When done searching, continue your reasoning. \\
\hline
\end{tabular}
\caption{Prompt Used for the Reasoning Assistant}
\label{tab:assistant_prompt}
\end{table*}

\begin{table*}[t]
\centering
\begin{tabular}{|p{0.95\linewidth}|}
\hline
Please extract content from the provided text that is useful for answering the given query, specifically with respect to the listed subjects, actions, temporal markers, degree modifiers. \\

Input: \\
- List of subjects, actions, temporal markers, degree modifiers: \\
- Text: ...\\
- Query: ... \\

Instructions: \\
- For each item in the list of subjects, actions, temporal markers or degree descriptions, extract raw, verbatim content from the text that is directly relevant to the query. \\
- Format each extracted piece as: \\
\quad \texttt{subjects: [exact content from the text about the subjects]} \\
\quad \texttt{actions: [exact content from the text describing an action involving the verb]} \\
\quad \texttt{temporal markers: [exact content from the text indicating when an event occurred or the time duration]} \\
\quad \texttt{degree modifiers: [exact content from the text indicating the characteristics of the actions and subjects.]} \\
- Do not paraphrase, summarize, or use your own words. Use only direct excerpts or minimally truncated phrases that preserve original wording. \\
- Each subjects/actions/temporal markers/degree modifiers–content pair must appear on a separate line. \\
- Only include content that provides diverse and query-relevant information. \\
- If no relevant content exists in the text for a given entity, verb, or time expression with respect to the query, omit it entirely. \\
\hline
\end{tabular}
\caption{Prompt for Memory Fragments Growth}
\label{tab:extraction_prompt}
\end{table*}

\begin{table*}[t]
\centering
\begin{tabular}{|p{0.95\linewidth}|}
\hline
Answer the question based on the given system memory of reasoning. Just give the answer and do not output other information. \\
You should provide your final answer in the format \textbackslash boxed\{YOUR\_ANSWER\}. \\

If the answer is in the context, maintain the illustrations (e.g., examples and specific phrasings) present in the context when formulating the answer. \\

System memory: ... \\
Question:... \\

Generate an accurate answer based solely on the provided information. \\
\hline
\end{tabular}
\caption{Prompt for Answering Questions Based on Given Context}
\label{tab:qa_prompt}
\end{table*}

\begin{table*}[t]
\centering
\begin{tabular}{|p{0.95\linewidth}|}
\hline
Prompt Template for Multiple Choice Questions \\
Please read the provided contexts and answer the question below. \\
\textless text\textgreater \{Contexts\} \textless /text\textgreater \\
What is the correct answer to this question: \{query\} \\
Choices: \\
(A) \{item['choice\_A']\} \\
(B) \{item['choice\_B']\} \\
(C) \{item['choice\_C']\} \\
(D) \{item['choice\_D']\} \\
Answer the question based on the given context. Just give the answer and do not output other information. Format your response as follows: "The correct answer is (insert answer here)". \\
\hline
\end{tabular}
\caption{Prompt Template for Multiple Choice Questions in LongBench v2}
\label{tab:longbench_v2_prompt}
\end{table*}


\end{document}